\documentclass{article}

\usepackage{arxiv}

\usepackage[utf8]{inputenc} % allow utf-8 input
\usepackage[T1]{fontenc}    % use 8-bit T1 fonts
\usepackage{hyperref}       % hyperlinks
\usepackage{url}            % simple URL typesetting
\usepackage{booktabs}       % professional-quality tables
\usepackage{amsfonts}       % blackboard math symbols
\usepackage{amssymb}        % needed for \leqslant
\usepackage{amsmath}        % needed for cases
\usepackage{nicefrac}       % compact symbols for 1/2, etc.
\usepackage{microtype}      % microtypography
\usepackage{lipsum}		% Can be removed after putting your text content
\usepackage{graphicx}
\usepackage[numbers]{natbib}
\usepackage{doi}
\usepackage{wrapfig}
\usepackage{tikz-cd}
\usepackage{xcolor}

\title{Antiassociative algebra in R: introducing the {\tt evitaicossa} package}

\author{ \href{https://orcid.org/0000-0001-5982-0415}{\includegraphics[width=0.03\textwidth]{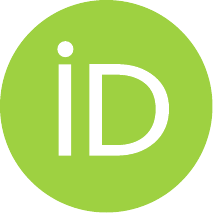}\hspace{1mm}Robin K. S.~Hankin}\thanks{\href{https://www.stir.ac.uk/people/1966824}{work};  
\href{https://www.youtube.com/watch?v=JzCX3FqDIOc&list=PL9_n3Tqzq9iWtgD8POJFdnVUCZ_zw6OiB&ab_channel=TrinTragulaGeneralRelativity}{play}} \\
University of Stirling\\
	\texttt{hankin.robin@gmail.com} \\
}

\hypersetup{
pdftitle={Antiassociative algebra in R},
pdfsubject={q-bio.NC, q-bio.QM},
pdfauthor={Robin K. S.~Hankin},
pdfkeywords={Antiassociative algebra}
}

\usepackage{Sweave}
\begin{document}
\maketitle

\setlength{\intextsep}{0pt}
\begin{wrapfigure}{r}{0.2\textwidth}
  \begin{center}
\includegraphics[width=1in]{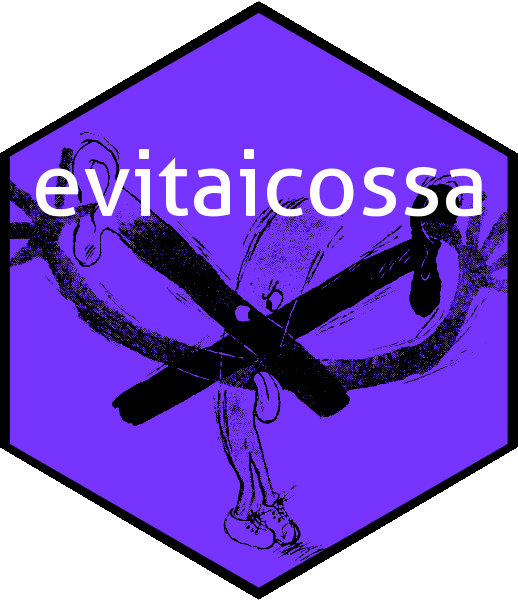}
  \end{center}
\end{wrapfigure}

\begin{abstract}

  In this short article I introduce the {\tt evitaicossa} package
  which provides functionality for antiassociative algebras in the R
  programming language; it is available on CRAN
  at\\ \url{https://CRAN.R-project.org/package=evitaicossa}.

\end{abstract}

\section{Introduction}

Here I introduce the {\tt evitaicossa} R package for antiassociative
algebras.  An {\em algebra} is a vector space in which the vectors
possess a well-behaved bilinear product.  Formally, a vector space is
a set $V$ of vectors which form an Abelian group under addition and
also satisfy the following axioms:

\begin{itemize}
\item Compatibility, $a(b\mathbf{v})=(ab)\mathbf{v}$
\item Identity, $1\mathbf{v}=\mathbf{v}$
\item Distributivity of vector addition, $a(\mathbf{u}+\mathbf{v}) =
  a\mathbf{u}+a\mathbf{v}$
\item Distributivity of field addition, $(a+b)\mathbf{u} =
  a\mathbf{u}+b\mathbf{u}$
\end{itemize}

Above, $\mathbf{u},\mathbf{v}\in V$ are vectors, $a,b$ are scalars
[here the real numbers], and $1$ is the multiplicative identity.  We
also require a bilinear vector product, mapping pairs of vectors to
vectors; vector multiplication is denoted using juxtaposition, as in
$\mathbf{u}\mathbf{v}$, which satisfies the following axioms:

\begin{itemize}
\item Right distributivity, $(\mathbf{u}+\mathbf{v})\mathbf{w} =
  \mathbf{u}\mathbf{w}+ \mathbf{u}\mathbf{w}$
\item Left distributivity, $\mathbf{w}(\mathbf{u}+\mathbf{v})  =
  \mathbf{w}\mathbf{u}+ \mathbf{w}\mathbf{v}$
\item Compatibility, $(a\mathbf{u})(b\mathbf{v})=(ab)(\mathbf{u}\mathbf{v})$
\end{itemize}

Note the absence of commutativity and associativity.  Associative
algebras seem to be the most common, and examples would include
multivariate polynomials
\citep{hankin2022_mvp_arxiv,hankin2022_spray_arxiv}, Clifford algebras
\citep{hankin2022_clifford_arxiv}, Weyl algebras
\citep{hankin2022_weyl_arxiv}, and free algebras
\citep{hankin2022_freealg_arxiv}.  Non-associative algebras would include
the octonions \citep{hankin2006_onion} and Jordan algebras
\citep{hankin2023_jordan_arxiv}.  Here I consider {\em antiassociative}
algebras in which the usual associativity relation
$\mathbf{u}(\mathbf{v}\mathbf{w})=(\mathbf{u}\mathbf{v})\mathbf{w}$ is
replaced by the relation
$\mathbf{u}(\mathbf{v}\mathbf{w})=-(\mathbf{u}\mathbf{v})\mathbf{w}$.

\section{Antiassociative algebras}

Algebras satisfying
$\mathbf{u}(\mathbf{v}\mathbf{w})=-(\mathbf{u}\mathbf{v})\mathbf{w}$
exhibit some startling behaviour.  Firstly, in the vector space there
are no scalars except for $0\in\mathbb{R}$.  Proof: for any
$x\in\mathbb{R}$, we have $x^3=x(xx)=-(xx)x=-x^3$; thus $x^3=-x^3$, so
$x=0$.  Secondly, antiassociative algebras are nilpotent of order 4:

$$
(\mathbf{a}\mathbf{b})(\mathbf{c}\mathbf{d}) =
-\mathbf{a}(\mathbf{b}(\mathbf{c}\mathbf{d})) =
\mathbf{a}((\mathbf{b}\mathbf{c})\mathbf{d}) =
-(\mathbf{a}(\mathbf{b}\mathbf{c}))\mathbf{d} =
((\mathbf{a}\mathbf{b})\mathbf{c})\mathbf{d} =
-(\mathbf{a}\mathbf{b})(\mathbf{c}\mathbf{d})
$$

We see that
$(\mathbf{a}\mathbf{b})(\mathbf{c}\mathbf{d})=-(\mathbf{a}\mathbf{b})(\mathbf{c}\mathbf{d})$
so $\mathbf{a}\mathbf{b}\mathbf{c}\mathbf{d}$ (however bracketed) must
be zero.

\subsection{The free antiassociative algebra}

We consider vector spaces generated by a finite alphabet of symbols
$\mathbf{x}_1,\ldots,\mathbf{x}_n$.  These will be denoted generally
by a single letter, as in $\mathbf{a},\mathbf{b},\ldots,\mathbf{z}$.
We now consider the algebra spanned by products of linear combinations
of these symbols, subject only to the axioms of an algebra [and the
antiassociative relation
$\mathbf{u}(\mathbf{v}\mathbf{w})=-(\mathbf{u}\mathbf{v})\mathbf{w}$].
Given an alphabet $\mathbf{x}_1,\ldots,\mathbf{x}_n$, the general form
of an element of an antiassociative algebra will be

$$
\sum_{i}\alpha_i\mathbf{x}_i +
\sum_{i,j}\alpha_{ij}\mathbf{x}_i\mathbf{x}_j+
\sum_{i,j,k}\alpha_{ijk}(\mathbf{x}_i\mathbf{x}_j)\mathbf{x}_k
$$

(see \citep{remm2024_arxiv} for a proof, but note that she uses
$\mathbf{x}_i(\mathbf{x}_j\mathbf{x}_k)$ rather than
$(\mathbf{x}_i\mathbf{x}_j)\mathbf{x}_k$ for the triple products; a
brief discussion is given in the appendix).  In the package, the
components of the first term $\sum_{i}\alpha_i\mathbf{x}_i$ are known
as ``single-symbol" terms [$\mathbf{x}_i$] and coefficients
[$\alpha_i$] respectively.  Similarly, the components of
$\sum_{i,j}\alpha_{ij}\mathbf{x}_i\mathbf{x}_j$ are known as the
``double-symbol" terms and coefficients; and the components of
$\sum_{i,j,k}\alpha_{ijk}(\mathbf{x}_i\mathbf{x}_j)\mathbf{x}_k$ are
the ``triple-symbol" terms and coefficients.

Addition is performed elementwise among the single-, double-, and
triple- components; the result is the (formal) composition of the
three results.  Given

$$A=
\sum_{i}\alpha_i\mathbf{x}_i +
\sum_{i,j}\alpha_{ij}\mathbf{x}_i\mathbf{x}_j+
\sum_{i,j,k}\alpha_{ijk}(\mathbf{x}_i\mathbf{x}_j)\mathbf{x}_k
$$

$$B=
\sum_{i}\beta_i\mathbf{x}_i +
\sum_{i,j}\beta_{ij}\mathbf{x}_i\mathbf{x}_j+
\sum_{i,j,k}\beta_{ijk}(\mathbf{x}_i\mathbf{x}_j)\mathbf{x}_k
$$

(where the sums run from $1$ to $n$), we define the sum $A+B$ to be

$$
\sum_{i}(\alpha_i+\beta_i)\mathbf{x}_i +
\sum_{i,j}(\alpha_{ij}+\beta_{ij})\mathbf{x}_i\mathbf{x}_j+
\sum_{i,j,k}(\alpha_{ijk}+\beta_{ijk})(\mathbf{x}_i\mathbf{x}_j)\mathbf{x}_k
$$

Multiplication is slightly more involved.  We define the product $AB$
to be

$$
\sum_{i,j}\alpha_i\beta_{ij}\mathbf{x}_i\mathbf{x}_j
+\sum_{i,j,k}\alpha_{ij}\beta_{k}(\mathbf{x}_i\mathbf{x}_j)\mathbf{x}_k
-\sum_{i,j,k}\alpha_i\beta_{jk}(\mathbf{x}_i\mathbf{x}_j)\mathbf{x}_k.
$$

The minus sign in front of the third term embodies antiassociativity.

\section{The {\tt evitaicossa} package}

The {\tt evitaicossa} package implements these relations in the
context of an R-centric suite of software.  I give some examples of
the package in use.  A good place to start is function {\tt raaa()},
which returns a simple random element of the free antiassociative
algebra:

\begin{Schunk}
\begin{Sinput}
> raaa()
\end{Sinput}
\begin{Soutput}
free antiassociative algebra element:
+1a +3b +2d +3a.b +1c.b +1c.c +1(a.b)a +1(b.b)c +1(b.c)a
\end{Soutput}
\end{Schunk}

(the default alphabet for this command is
$\left\lbrace\mathbf{a},\mathbf{b},\mathbf{c},\mathbf{d}\right\rbrace$).
We see the print method for the package which shows some of the
structure of the object.  This one has some single-symbol elements,
some double-symbol and some triple-symbol elements.

It is possible to create elements using the {\tt aaa()} or {\tt as.aaa()}
functions:

\begin{Schunk}
\begin{Sinput}
> x  <- as.aaa(c("p","q","r"))
> x1 <- aaa(s1 = c("p","r","x"),c(-1,5,6))
> y <- aaa(d1 = letters[1:3],d2 = c("foo","bar","baz"),dc=1:3)
> z <- aaa(
+ 	t1 = c("bar","bar","bar"),
+ 	t2 = c("q","r","s"),
+ 	t3 = c("foo","foo","bar"),
+ 	tc = 5:7)
\end{Sinput}
\end{Schunk}

And then apply arithmetic operations to these objects:

\begin{Schunk}
\begin{Sinput}
> x
\end{Sinput}
\begin{Soutput}
free antiassociative algebra element:
+1p +1q +1r
\end{Soutput}
\begin{Sinput}
> x1
\end{Sinput}
\begin{Soutput}
free antiassociative algebra element:
-1p +5r +6x
\end{Soutput}
\begin{Sinput}
> x+x1
\end{Sinput}
\begin{Soutput}
free antiassociative algebra element:
+1q +6r +6x
\end{Soutput}
\end{Schunk}

(above, note the cancellation in {\tt x+x1}).  Multiplication is also
implemented (package idiom is to use an asterisk):

\begin{Schunk}
\begin{Sinput}
> x*(x1+y)
\end{Sinput}
\begin{Soutput}
free antiassociative algebra element:
-1p.p +5p.r +6p.x -1q.p +5q.r +6q.x -1r.p +5r.r +6r.x -1(p.a)foo -2(p.b)bar
-3(p.c)baz -1(q.a)foo -2(q.b)bar -3(q.c)baz -1(r.a)foo -2(r.b)bar -3(r.c)baz
\end{Soutput}
\end{Schunk}

Check:

\begin{Schunk}
\begin{Sinput}
> x*(x1+y) == x*x1 + x*y
\end{Sinput}
\begin{Soutput}
[1] TRUE
\end{Soutput}
\end{Schunk}

We end with a remarkable identity:

$$
(\mathbf{a}+\mathbf{a}\mathbf{x})(\mathbf{b} + \mathbf{x}\mathbf{b})=\mathbf{a}\mathbf{b}
$$

Numerically:

\begin{Schunk}
\begin{Sinput}
> a <- raaa()
> b <- raaa()
> x <- raaa()
> (a+a*x)*(b+x*b) == a*b
\end{Sinput}
\begin{Soutput}
[1] TRUE
\end{Soutput}
\end{Schunk}

\section{Extract and replace methods}

Because of the tripartite nature of antiassociative algebra, the
package provides three families of extraction methods: {\tt single()},
{\tt double()} and {\tt triple()}, which return the different
components of an object:

\begin{Schunk}
\begin{Sinput}
> a
\end{Sinput}
\begin{Soutput}
free antiassociative algebra element:
+4a +2b +2a.a +2c.c +2d.d +2(b.d)d +1(c.d)a +4(d.b)c
\end{Soutput}
\begin{Sinput}
> single(a)
\end{Sinput}
\begin{Soutput}
free antiassociative algebra element:
+4a +2b
\end{Soutput}
\begin{Sinput}
> double(a)
\end{Sinput}
\begin{Soutput}
free antiassociative algebra element:
+2a.a +2c.c +2d.d
\end{Soutput}
\begin{Sinput}
> triple(a)
\end{Sinput}
\begin{Soutput}
free antiassociative algebra element:
+2(b.d)d +1(c.d)a +4(d.b)c
\end{Soutput}
\end{Schunk}

The corresponding replacement methods are also implemented:

\begin{Schunk}
\begin{Sinput}
> single(a) <- 0
> a
\end{Sinput}
\begin{Soutput}
free antiassociative algebra element:
+2a.a +2c.c +2d.d +2(b.d)d +1(c.d)a +4(d.b)c
\end{Soutput}
\begin{Sinput}
> double(a) <- double(b) * 1000
> a
\end{Sinput}
\begin{Soutput}
free antiassociative algebra element:
+1000b.d +2000c.b +2000c.d +2(b.d)d +1(c.d)a +4(d.b)c
\end{Soutput}
\end{Schunk}

Square bracket extraction and replacement is also implemented:

\begin{Schunk}
\begin{Sinput}
> (a <- raaa(s=5))
\end{Sinput}
\begin{Soutput}
free antiassociative algebra element:
+4a +3b +6c +3b.c +1b.d +2c.c +3d.a +2d.c +2(a.d)d +3(b.a)a +4(b.b)c +1(c.d)d
+3(d.c)b
\end{Soutput}
\begin{Sinput}
> a[s1=c("c","e"),t1="c",t2="d",t3="d"]
\end{Sinput}
\begin{Soutput}
free antiassociative algebra element:
+6c +1(c.d)d
\end{Soutput}
\end{Schunk}

Above we pass named arguments ({\tt s1} {\em et seq.}) and the
appropriate {\tt aaa} object is returned.  Zero coeffients are
discarded.  This mode also implements replacement methods:

\begin{Schunk}
\begin{Sinput}
> (a <- raaa(s=5))
\end{Sinput}
\begin{Soutput}
free antiassociative algebra element:
+8a +2b +2c +1a.b +4a.c +3b.c +4c.c +2(a.c)c +4(b.c)a +4(b.c)d +1(c.d)d
+3(d.c)a
\end{Soutput}
\begin{Sinput}
> a[s1="a",d1=c("c","w"),d2=c("d","w")] <- 888
> a
\end{Sinput}
\begin{Soutput}
free antiassociative algebra element:
+888a +2b +2c +1a.b +4a.c +3b.c +4c.c +888c.d +888w.w +2(a.c)c +4(b.c)a
+4(b.c)d +1(c.d)d +3(d.c)a
\end{Soutput}
\end{Schunk}

The other square bracket method is to pass an (unnamed) character
vector:

\begin{Schunk}
\begin{Sinput}
> (a <- raaa())
\end{Sinput}
\begin{Soutput}
free antiassociative algebra element:
+1a +2c +4d +3b.a +3b.b +3b.c +2(b.c)a +4(c.a)d +4(c.d)b
\end{Soutput}
\end{Schunk}

\section{Note on {\tt disordR} discipline}

If we try to access the symbols or coefficients of an {\tt aaa} object
[functions {\tt s1()} and {\tt sc()} respectively], we get a {\tt
  disord} object \citep{hankin2022_disordR_arxiv}.  Suppose we wish to
extract the single-symbol terms and the single-symbol coefficients:

\begin{Schunk}
\begin{Sinput}
> x
\end{Sinput}
\begin{Soutput}
free antiassociative algebra element:
+3b +4c +1d +3a.c +2a.d +4d.a +1(a.a)d +1(a.b)d +3(d.d)a
\end{Soutput}
\begin{Sinput}
> s1(x)
\end{Sinput}
\begin{Soutput}
A disord object with hash b3ead18cdb285c1bd26ea0d9faf1dfcff2debd81 and elements
[1] "b" "c" "d"
(in some order)
\end{Soutput}
\begin{Sinput}
> sc(x)
\end{Sinput}
\begin{Soutput}
A disord object with hash b3ead18cdb285c1bd26ea0d9faf1dfcff2debd81 and elements
[1] 3 4 1
(in some order)
\end{Soutput}
\end{Schunk}

See how the hash codes of the symbols and coeffients match.  However,
the double-symbol terms and coefficients, while internally matching,
differ from the single-symbol stuff:

\begin{Schunk}
\begin{Sinput}
> list(d1(x),d2(x),dc(x))
\end{Sinput}
\begin{Soutput}
[[1]]
A disord object with hash eeb24e0f65a9b27e547fc0c7be14e7b33be06cd0 and elements
[1] "a" "a" "d"
(in some order)

[[2]]
A disord object with hash eeb24e0f65a9b27e547fc0c7be14e7b33be06cd0 and elements
[1] "c" "d" "a"
(in some order)

[[3]]
A disord object with hash eeb24e0f65a9b27e547fc0c7be14e7b33be06cd0 and elements
[1] 3 2 4
(in some order)
\end{Soutput}
\end{Schunk}

Above, see how the double-symbol terms and double-symbol coefficients
have consistent hashes, but do not match the single-symbol objects (or
indeed the triple-symbol objects).

\subsection{Matrix index extraction}

If square bracket extraction is given an index that is a matrix, it is
interpreted rowwise:

\begin{Schunk}
\begin{Sinput}
> l <- letters[1:3]
> (a <- aaa(s1=l,sc=1:3, d1=l,d2=rev(l),dc=3:1,t1=l,t2=l,t3=rev(l),tc=1:3))
\end{Sinput}
\begin{Soutput}
free antiassociative algebra element:
+1a +2b +3c +3a.c +2b.b +1c.a +1(a.a)c +2(b.b)b +3(c.c)a
\end{Soutput}
\begin{Sinput}
> a[cbind(l,l)]
\end{Sinput}
\begin{Soutput}
free antiassociative algebra element:
+2b.b
\end{Soutput}
\begin{Sinput}
> a[cbind(rev(l),l,l)] <- 88
> a
\end{Sinput}
\begin{Soutput}
free antiassociative algebra element:
+1a +2b +3c +3a.c +2b.b +1c.a +1(a.a)c +88(a.c)c +88(b.b)b +88(c.a)a +3(c.c)a
\end{Soutput}
\end{Schunk}

\section{Note on generalized antiassociativity}

We may generalize antiassociativity to
$\mathbf{a}(\mathbf{b}\mathbf{c})=k(\mathbf{a}\mathbf{b})\mathbf{c}$.
Thus associativity is recovered if $k=1$ and antiassociativity if
$k=-1$.  Then the nilpotence argument becomes:

$$
(\mathbf{a}\mathbf{b})(\mathbf{c}\mathbf{d}) =
k^{-1}\mathbf{a}(\mathbf{b}(\mathbf{c}\mathbf{d})) =
\mathbf{a}((\mathbf{b}\mathbf{c})\mathbf{d}) =
k(\mathbf{a}(\mathbf{b}\mathbf{c}))\mathbf{d} =
k^2((\mathbf{a}\mathbf{b})\mathbf{c})\mathbf{d} =
k(\mathbf{a}\mathbf{b})(\mathbf{c}\mathbf{d})
$$

The value of $k$ may be set at compile-time by editing file
{\tt src/anti.h}.  The line in question reads:

\begin{verbatim}
#define K -1 // a(bc) == K(ab)c
\end{verbatim}

but it is possible to change the value of {\tt K}.  Note that this
will cause {\tt test\_aac.R}, one of the {\tt testthat} suite, to fail
{\tt R CMD check}.

\appendix

As noted above, \cite{remm2024_arxiv} uses
$\mathbf{x}_i(\mathbf{x}_j\mathbf{x}_k)$ rather than
$(\mathbf{x}_i\mathbf{x}_j)\mathbf{x}_k$ for the triple products.  I
chose the latter because R idiom for multiplication is left
associative:

\begin{Schunk}
\begin{Sinput}
> x <- 3
> class(x) <- "foo"
> `*.foo` <- function(x,y){x + y + x}
> print.foo <- function(x){print(unclass(x))}
> c(`(x*x)*x` = (x*x)*x,  `x*(x*x)` = x*(x*x),  `x*x*x` = x*x*x)
\end{Sinput}
\begin{Soutput}
(x*x)*x x*(x*x)   x*x*x 
     21      15      21 
\end{Soutput}
\end{Schunk}

Above we see that {\tt x*x*x} is interpreted as {\tt (x*x)*x}, which
is why the sign convention in the package was adopted.

\bibliographystyle{apalike}
\bibliography{evitaicossa}

\end{document}